\documentclass[12pt]{article}
\usepackage{amssymb,amsmath}

\usepackage{epsfig}
\usepackage{cite}
\usepackage{listings}

\makeatletter \@addtoreset{equation}{section} \makeatother

\def\p{\partial}
\def\tr{{\rm tr}}
\def\Tr{{\rm Tr}}

\def\a{\alpha}\def\g{\gamma}
\def\d{\delta}\def\e{\epsilon}

\def\l{\lambda}
\def\m{\mu}\def\n{\nu}
\def\r{\rho}\def\s{\sigma}
\def\t{\tau}
\def\w{\omega}\def\G{\Gamma}

\newcommand{\be}{\begin{eqnarray}}
\newcommand{\ee}{\end{eqnarray}}
\newcommand{\nn}{\nonumber}

\begin{document}

\begin{titlepage}
\vfill
\begin{flushright}
{\tt\normalsize KIAS-P12064}\\
\end{flushright}
\vfill
\begin{center}
{\large\bf  Supersymmetric M5 Brane Theories   on $\rm R \times  CP^2$ }

\vfill
{\large  Hee-Cheol Kim and Kimyeong Lee \footnote{\tt e-mail:
 heecheol1@gmail.com, klee@kias.re.kr,  }}

{ \it
\vskip 5mm
Korea Institute for Advanced Study, Seoul 130-722, Korea
}
\vfill
\end{center}

\begin{abstract}
\noindent   We propose 4 and 12 supersymmetric conformal   Yang-Mills-Chern-Simons theories on $\mathrm{ R\times CP^2}$ as multiple representations of  the theory on M5 branes. These theories  are obtained by    twisted $\mathrm{Z}_k$ modding  and dimensional reduction of the 6d (2,0) superconformal field theory on $\mathrm{R\times S^5}$ and    have a  discrete coupling constant $\frac{1}{g^2_{YM}} =\frac{k}{4\pi^2}$ with natural number $k$. Instantons in these theories are expected to   represent  the Kaluza-Klein modes.  For the $k=1,2$ cases, we argue that the number of supersymmetries in our theories should be enhanced to $32$ and $16$, respectively. For the $k=3$ case, only the 4 supersymmetric theory gets the supersymmetric enhancement to $8$.   For  the 4 supersymmetric case, the vacuum structure becomes more complicated as there are degenerate supersymmetric vacua characterized by   fuzzy spheres. We  calculate  the perturbative part of the $SU(N)$ gauge group Euclidean path integral for the index function at the symmetric phase of the 4 supersymmetric case and confirm it with the known half-BPS index.   From the similar twisted $Z_k$ modding of the $\mathrm{AdS_7\times S^4}$ geometry, we speculate that the $M$   region is for $k\lesssim N^{1/3}$ and the type IIA   region is $N^{1/3}\lesssim k \lesssim N$. When nonperturbative corrections are included, our theories are expected to produce the full index of the 6d (2,0) theory.
  \end{abstract}

\vfill
\end{titlepage}

\section{Introduction}

The physics of M5 branes~\cite{witten1,strominger} remains as one of great mysteries in M-theory~\cite{hull,witten1}. Some fundamental structures of the underlying 6d (2,0) superconformal theory   are
not yet known. One  promising approach is  to study the 5d maximally supersymmetric gauge theory whose instantons may provide all Kaluza-Klein physics of the circle-compactified 6d theory~\cite{douglas,lambert}. The study of a 1/4 BPS sector by  the index calculation in this setting has provided the exact results on the DLCQ limit of the 6d (2,0) theory~\cite{aharony,instanton}. However, one wants to have more divices to probe this 6d theory which allow, for example,  the calculation of the full index function on $\mathrm{S^1\times S^5}$.

In this work we propose  one such tool.   First we put  the 6d (2,0) theory on $\mathrm{R \times S^5}$. The five sphere $\rm S^5$ is a circle fibration over $\mathrm{CP^2}$,  and we mod out the theory by  $Z_k$ along this circle fiber   with some additional twisting along a $U(1)$ subgroup   of the $Sp(2)_R= SO(5)$  R-symmetry. This allows a consistent truncation of the 6d theory to  a 5d theory on $\mathrm{R\times  CP^2}$ with partially conserved supersymmetries. While we do not know the exact nonabelian 6d (2,0) theory, one can find this  5d nonabelian theory explicitly. The 5d theory  has both Yang-Mills     Chern-Simons terms  and the Myers term for the scalar field. The Chern-Simons term is not the standard 5d Chern-Simons term but is
of type  $JAdA$ where $J$ is the  K\"ahler form on $\rm CP^2 $.   This $Z_k$ modding and dimensional reduction lead  to the overall coupling constant $1/g_{YM}^2=k/4\pi^2 r$ with the $\rm S^5$ radius $r$, and so the 5d theories with $SU(N)$ gauge group have a weakly coupled regime with the small 't Hooft coupling constant $\l = N/k$ when $k\gg N$.  A different choice of twisting leads to a different 5d theory, even with different amount of supersymmetries.     Here we construct two such 5d theories with either 4 or 12 supersymmetries.

 The Killing spinors of $\rm S^5$ can be singlet or triplet under  the isometry group $SU(3)$ of $\rm CP^2 $. We will show that the   4 and 12 supersymmetric theories have   singlet  and   triplet Killing spinors, respectively. Our 5d theories  do  not appear in the standard classification of the super conformal field theories as  Poincare supersymmetry is partially broken here~\cite{nahm,kac}.
The supersymmetry on $\rm R\times CP^2$ inherits the original superconformal symmetry and  the eigenvalues of  our Hamiltonian   can be identified with  the conformal dimension of an  operator corresponding to the eigenstate  on $ \rm CP^2$.  As we have modded out some sector of the original theory, the quantum states of our theories for $k>1$ have less amount of the quantum states than the original 6d theory.

As instantons in the 5d maximally supersymmetric Yang-Mills theory obtained by the dimensional reduction of the 6d (2,0) theory on $\rm R^5\times S^1$ are supposed to provide  the Kaluza-Klein modes~\cite{rozali,Seiberg:97}, instantons  in our theories are assumed also to   provide  all of the KK physics.  Our theory has only one coupling constant which is discrete and quantized.  In addition we expect that for $k=1$  the amount of the  supersymmetries of our theories should be enhanced to 32 as there is no modding.  Especially,  our theories with $k=1$  could capture all the physics of the 6d (2,0) theory on $\rm R\times S^5$ and so become equivalent to the 6d (2,0) theory.  The eigenstates of the Hamiltonian of our 5d theories for $k=1$ with all non-perturbative effect included should be those of the 6d Hamiltonian on ${\rm R\times S^5}$. We will argue later on that for $k=2$ case  the number of supersymmetries of our 5d theories should be enhanced to 16. For $k=3$ case, only  the 4 supersymmetric theory gets the supersymmetric enhancement to $8$.  For $k\ge 4$ case we do not expect any enhancement.

The supersymmetry of our 5d theories is a  part of 6d superconformal symmetry and so allows    the definition of the superconformal indices in the 6d sense. Here we calculate the conformal index in the large $k$ or free theory limit for the 4 supersymmetric theory and found that it matches exactly to what is expected.

There are three possibilities for our theory in the ultraviolet region: (1) UV finite and complete, (2) UV non-finite but renormalizable, and (3) UV non-finite and non-renormalizable. Unlike 5d SYM on $\rm R^{1+4}$ which has the dimensionful coupling with no additional adjustable parameter, our theory has the unique discrete quantized coupling constant with Chern-Simons term  and also the weak coupling regime,  and so has a better chance to be UV finite.  It would be fascinating to  figure out whether this is the case.

The $\mathrm{AdS_7\times S^4}$  geometry for the large $N$ M5 branes is known~\cite{adscft} and    a  similar $Z_k$ modding of this geometry would lead to the geometry corresponding to our theories.  These geometries are unusual as the asymptotic geometry is not $\mathrm{AdS_6}$. We speculate that there are three regions of $k$: the M-theory region for $1\le k \lesssim  N^{1/3}$, the type IIA region for $N^{1/3}\lesssim k\lesssim N$, and the high curvature region for $N\lesssim k$. Such division is not concrete as the 11d circle radius, that is  the dilation field,  diverges at the boundary, which is the UV region of the field theory. These regions could be meaningful in the interior region of the $AdS_7$ space.

Our approach is  inspired in part by the ABJM theory on M2 branes which has $Z_k$ modding of the $SO(8)$ R-symmetry~\cite{abjm}. However, our $Z_k$ modding is acting on both the   space $\mathrm{S^5}$ and the $\rm R^2$ part of the scalar field space $\mathrm{R^5}$.  Our 5d theories are defined on a compact space  $\mathrm{ CP^2}$ with $SU(3)$ isometry instead of non-compact $R^{4}$.   

We note that 5d $JFA$ type Chern-Simons term has appeared in Ref.~\cite{ohlsson,kallen0} while their setting is different from ours. There is another work by one of us (HK) and Seok Kim where the index on M5 brane has been approached by the 5d Yang-Mills theory on $\mathrm S^5$~\cite{seok1}.  Not only a perturbative calculation on $\mathrm S^5$ is done explicitly there but also a conjecture on instanton part  has been provided. More relevant for the future work would be the index calculation on $\mathrm{S^1\times S^4}$ done recently for the 5d superconformal field theories~\cite{hee-cheol}. There are some related recent works~\cite{hosomichi0,kallen1,jafferis1}  on the supersymmetric theories on $\mathrm S^5$.

A Myers' term   appears  in our 5d theories. For 4 supersymmetric case with $SU(N)$ gauge group, one can have  degenerate vacua  which are characterized by supersymmetric fuzzy spheres and the partition of $N$. This classical degeneracy of vacua  could be regarded as a blow up of D4 world volume from $\rm CP^2$ to $\rm CP^2\times S^2$ and thus implies that   some D6 brane giant gravitons  contribute to the index.  Such possibility raises many interesting questions which we would leave   as a future problem.

All the fields in our 5d theories belong to the adjoint representation of the gauge group and the overall coupling constant is given as $1/g_{YM}^2= k/4\pi^2r$ with natural number $k$ and the $S^5$ radius $r$. These  5d theories,  obtained after the  $Z_k$ modding and the dimensional reduction, have the weak coupling limit in large $k$ and discrete coupling constants beside the dimensionful factor $r$. Our 5d theories could be ultraviolet complete when nonperturbative part is included. But our theories are not defined on flat $R^{1+4}$ with Lorentz symmetry and so the usual perturbative expansion in momentum space is not available.     The large $k$ limit is the weak coupling limit and the $k=1$ limit is the strong coupling limit. For the $SU(N)$ or $U(N)$ theory, there is also 't Hooft coupling constant $\l=N/k$.  For large 't Hooft coupling limit   the corresponding AdS  geometry is obtained by the $Z_k$ modding of the $\mathrm{AdS_7 \times S^4}$ and somewhat complicated as the boundary geometry is a $Z_k$ modding of the boundary geometry $\mathrm{R\times S^5\times S^4}$.

The index function for a conformal field theory is an important tool to explore the theory~\cite{kinney,romelsberger,seok0}.
The index function of the 6d (2,0) theory on $\mathrm{S^1\times S^5}$ is one of the major interests. The index for the U(1) theory on a single M5 has been done~\cite{Bhattacharyya2}. Our 5d theories have both perturbative parts and instanton parts. In this work, we restrict ourself to   just the perturbative part of 4-supersymmetric case and find it to match with the known 1/2 BPS index
on the single M5 brane~\cite{Bhattacharyya1}.

The outline of this work is as follows. In Sec.2 we start with the 6d abelian (2,0) theory and do the twisted $Z_k$ modding and the dimensional reduction to obtain some supersymmetric 5d Yang-Mills Chern-Simons theories on $\mathrm{R\times CP^2}$. In Sec.3  we explore the properties of these theories, including the spectrum of the abelian theory. In Sec.4 we introduce the index function and calculate it by the Euclidean path integral in the weak coupling limit. In Sec.5 we conclude with some remarks. In Appendices we include the properties of manifolds $ \rm S^5$ and $\mathrm{CP^2}$   and the Killing spinors on them.

 \section{5d Supersymmetric Theories on $\mathrm{ R\times CP^2}$}

Let us start with the 6d abelian (2,0) theory on $\rm R^{1+5}$ for the field $B_{MN}, \l, \phi_I (I=1,2,3,4,5)$. The 3-form field strength $H=dB$ should be selfdual $H={}^*\!\!\!~H$. We
start with the supersymmetric action with additional spectator field $H=-{}^*\!\!\!~H$ which does not get involved in the supersymmetric transformation~\cite{mons}. The bosonic part of the superconformal symmetry $OSp(2,6|2)$ is made of the $SO(2,6)$ conformal symmetry and $Sp(2)_R=SO(5)$ R-symmetry. The conformal dimensions of $H,\l,\phi_I$ are
$3,5/2,2$, respectively. One does the radial quantization of the theory and obtains the Minkowski time action on $\rm R\times S^5$.  The Cartan elements of the spatial rotation algebra $SU(4)=SO(6)$ are made of $j_1,j_2,j_3$ and the Cartan elements of the Lie algebra $Sp(2)_R=SO(5)$ are made of $R_1,R_2$.
  The $R_1$ rotates the scalar fields $\phi_1,\phi_2$ and $R_2$ rotates $\phi_4,\phi_5$. Spinor field $\l$ belongs to   ${\bf 4}$ of $SU(4)$ and ${\bf 4}$ of $Sp(2)_R$. Both the fermion field $\l$ and supercharge $Q$ transform identically under $SU(4)$ and $Sp(2)_R$.
In terms of roots $\pm e_i \pm e_j, (i,j=1,2,3)$ of $SO(6)$, the spinor representations ${\bf 4}$ and $\bar{\bf 4}$ of $SU(4)$ are given by the weights $(\pm e_1\pm   e_2\pm e_3)/2$ with odd and even numbers of minus signs, respectively.

Let us do the radial quantization of the (2,0)  theory on $\rm R\times S^5$. See the appendix A for the metrics on $\rm S^5$ and $\rm CP^2$.
The action on $R\times S^5$  is
\be  S = \int_{R\times S^5} d^6x\sqrt{g}  \ \Big\{ -\frac{1}{12} H_{MNP}H^{MNP} -\frac{i}{2}\bar \l \G^M\hat\nabla_M\l
- \frac12 \p_M \phi_I \p^M\phi_I - \frac{2}{r^2} \phi_I\phi_I \Big\} .\ee
Here $\hat\nabla_M$ is the   spinor covariant derivative on $\rm R\times S^5$. From now on we put the $\rm S^5$ radius $r$ to be unity for the simplicity. The supersymmetric transformation for the tensor multiplet is
\be \d B_{MN} &=&  -\bar \l \G_{MN} \e = - \bar\e \G_{MN}\l, \nn  \\
\d \phi_I &=&  -\bar\l \r_I\e = +\bar\e \r_I \l, \nn \\
\d \l &=& +\frac{i}{6}  H_{MNP} \G^{MNP}\e +i \p_M\phi_I \G^M\r_I\e  - 2  \phi_I \r_I
 \tilde\e ,\nn \\
\d\bar\l &=& -\frac{i}{6}   H_{MNP}    \bar\e \G^{MNP} +i \p_M\phi_I \bar\e \G^M \rho_I -2\bar{\tilde\e}     \r_I \phi_I.
\ee
The Killing spinors $\e$ should satisfy
\be \hat\nabla_M \e  =   \frac{i}{2r} \G_M \tilde \e , \ \ \G^M\hat\nabla_M \tilde \e =  2i \e, \ee
which can be partially solved by $\tilde \e = \pm \G_0\e $.

Note that
\be H_{MNP}\G^{MNP}\e = \frac12 (H_{MNP} + ^*\!\!H_{MNP})\G^{MNP} \e, \ee
where
\be ^*\!  H_{MNP}=\frac16 \e_{MNPQRS}H^{QRS} , \ \e_{0123456}=-1. \ee
Only the selfdual part $H=^*\!\! H$ appears in the supersymmetry transformation.
Thus the anti-selfdual part of the field strength transform as
\begin{align}  \d (H_{MNP}- ^*\!\!H_{MNP}) & =i\bar \e \G_{MNP}\G^Q \p_Q \l,
 \end{align}
which vanishes on-shell.

The metric for the five sphere is
\be ds^2_{S^5} = ds^2_{\mathtt{CP}^2}  + (dy+V)^2 ,\ee
where $ y \sim y+ 2\pi$. The K\"ahler form $J$ is given by
\be J = \frac12 dV  . \ee
We want to a $Z_k$ modding of the 6d (2,0) theory along the fiber direction with identification
\be y\sim y + \frac{2\pi}{k}.   \ee

The Killing spinors on $\rm S^5$ as shown in appendix B have nontrivial $y$-dependence and the above $Z_k$ modding would remove them unless one introduces an additional twisting along some direction of $ Sp(2)_R=SO(5)$ R-symmetry.  Let us consider the plus sign case  $\tilde \e = +\G_0\e$. With the notation for the eigenspinors      $\g^{12}\e^{s_1s_2} = is_1\e^{s_1s_2}, \g^{34}\e^{s_1s_2}= is_2\e^{s_1s_2} $, we group  the 16 Killing spinors $\e_+$ to the $4$ and $12$ spinors.
The first group of the Killing spinors  is made of
\be ({\bf I}) \ \e_+ \sim e^{-\frac{i}{2}t +\frac{3i}{2}y} \e_0^{++}, \ee
with   constant spinors $\e_0^{++}$ which form a singlet of $SU(3)$ isometry of $\mathrm{CP^2}$ and the   fundamental representation of $Sp(2)_R=SO(5)$.  The second group of the  three independent  Killing spinors is made of  
\be ({\bf II}) \ \e_+ \sim e^{-\frac{i}{2}t -\frac{i}{2}y} ( \e_1^{+-},   \e_2^{-+} ,  \e_3^{--} ) , \ee
where these are   complicated $\rm CP^2$-dependent matrix linear combinations of three constant spinors. They form a triplet of $SU(3)$ isometry of $\mathrm{CP^2}$ and the  fundamental representation of $Sp(2)_R=SO(5)$.  The exact form is not important here.

We want to cancel the $y$-dependence of the spinor parameter by introducing a twisting of the spinor parameter along the $R$-symmetry direction. There are many inequivalent and  less supersymmetric choices. Here we focus on 
two choices for the simplicity and   twist both spinor and scalar fields   to be consistent with the supersymmetric transformation.

The  first choice is to introduce new variables
\be ({\bf I })\ \ \e_{old}=e^{-\frac{3\r_{45}}{2}y }\e_{new}, \ \l_{old}=e^{-\frac{3\r_{45}}{2}y}\l_{new}, \ (\phi_4+i\phi_5)_{old}= e^{+3iy } (\phi_4+i\phi_5)_{new}. \ee
This change of variables leads to
\be ({\bf I }) \ \ \p_y\rightarrow \p_y +3i R_2  \ee
on new variables. Here $R_2$ is one of the Cartans of $Sp(2)_R=SO(5)$ R-symmetry. The corresponding $U(1)_{R_2}$ transformation is      $\phi_4+i\phi_5  \rightarrow e^{i\a} (\phi_4+i\phi_5)$ and $\l \rightarrow e^{-\frac{\r_{45}}{2} \a}\l $.
The new spinor parameter has the $y$-dependence  as $e^{3(i + \r_{45})y/2} \e_{0}^{++}$, and so we choose the constant spinor to satisfy $\r_{45}\e_0= -i\e_0$ to remove  the $y$-dependence. This supersymmetry would survive under the $Z_k$ modding.
The possible $R_1$-charge  of allowed $\e_+$ spinors is  the eigenvalues $\pm i $ of $\r_{12}$.   As $\e_-$ is a complex conjugate, there would be four surviving supersymmetries in the first case after $Z_k$ modding. 

The second one is to introduce new variables so that
\be ({\bf II}) \ \ \ \e_{old} = e^{+\frac{\r_{45}}{2} y} \e_{new}, \ \l_{old} =e^{+\frac{\r_{45}}{2}y}\l_{new} , \ (\phi_4+i\phi_5)_{old} = e^{-iy} (\phi_4+i\phi_5)_{new}. \ee
This leads to the change of
the derivative $\p_y$ on new fields to
\be ({\bf II}) \ \ \p_y \rightarrow \p_y - i R_2.  \ee
 The $y$-dependence of the Killing spinors $\e_+$ can be removed once  $\r_{45}\e_+ = -i\e_+$. These two triplets of  Killing spinors $\e_+$ with eigenvalue of $\r_{12}=\pm i$ would survive under the $Z_k$ modding and so the resulting theory would have 12 supersymmetries.

The $Z_k$ modding of the new spinor and scalar fields is given 
to be
\begin{align} ({\bf I})\ & \l(y)_{old}\sim  e^{ -\frac{3\pi\r_{45}}{k}}\l(y+\frac{2\pi}{k})_{old},& (\phi_4+i\phi_5) (y)_{old} \sim  e^{+\frac{6\pi i}{k}}(\phi_4+i\phi_5)(y+\frac{2\pi}{k})_{old} & ,\nn \\
({\bf II})\ & \l(y)_{old}\sim  e^{+\frac{ \pi\r_{45}}{ k}}\l(y+\frac{2\pi}{k})_{old}, & (\phi_4+i\phi_5)(y)_{old}\sim  e^{-\frac{2\pi i}{k}}(\phi_4+i\phi_5)(y+\frac{2\pi}{k})_{old} & .
\end{align}
Such consistent $Z_k$ modding of the   6d (2,0) theory reduces the degrees of freedom and the number of supersymmetries. Still we do not know the exact form of the resulting 6d theory.

Let us now do the dimensional reduction of the 6d (2,0) abelian theory to 5d by requiring the new field variables to be independent of $y$. Then the $y$-independent    new spinor and scalar fields are invariant under the $Z_k$ modding and so are allowed. For the two-form tensor field, we can either do the dimensional reduction with identification $H_{\m\n 5}\sim F_{\m\n}$, or find the gauge kinetic term by the supersymmetry completion. 
 As we found the $y$ independent Killing spinors for the superconformal transformation, the supersymmetric transformation under these Killing spinors will not introduce additional $y$ dependence between the fields.  This $Z_k$ modding in large $k$ limit would shrink the circle fiber size relative to $\rm R\times CP^2$ size   and  so the theory would become more close to the 5d theory.

We first consider only the action for the scalar and spinor fields and  then fix the  the gauge kinetic term to complete the supersymmetry for the abelian case. Then one generalizes the theory to the nonabelian case. Only subtlety is  the right normalization for the coupling constant. For both cases we argue in the next section that   the 5d gauge coupling constant is given by
\be \frac{1}{g^2_{YM}} = \frac{k}{4\pi^2 r}, \ee
where $r$ is the radius of the $S^5$ sphere and is regard as unity as it is only length scale of the theory.
The theory becomes weakly coupled in large $k$ limit. As the fields are in the adjoint representation of the gauge group, one expects the presence of 't Hooft coupling constant
\be  \frac{N}{k} \ee
for $U(N)$ gauge theory.

To be more concrete, we use the four-component notation for the 5d  spinors as given in appendix A.  The symplectic reality condition becomes
\be \l = BC\l^*, \ \e=BC\e^*. \ee
The Killing spinor equation  for the $y$-independent new spinor parameter $\e_{new}$  becomes
\be
&  \p_t\e =\frac{i}{2} \g_0\tilde \e,  \   D_m \e = - \frac{i}{2} J_{mn}\g^n \e+ \frac{i}{2} \g_m  \tilde \e ,
\ee
where $m=1,2,3,4$. These spinor variables satisfy additional conditions for two cases:
\begin{align}
  ({\bf I}) & \ \    \r_{45}\e_+= -i\e_+, \  \  D_m = \nabla_m +\frac{3\r_{45}}{2} V_m ,\ \  \ \  \tilde \e = -\Big[3\r_{45}+ \frac12  J_{mn}\g^{mn}\Big]\e,
   \nn \\
  ({\bf II}) &  \ \   \r_{45}\e_+=-i\e_+,  \ \  D_m = \nabla_m -\frac{\r_{45}}{2} V_m, \ \ \  \ \  \tilde \e = \Big[\r_{45}-\frac12 J_{mn}\g^{mn}\Big] \e      ,
\end{align}
where $\nabla_m$ is the spinor covariant derivative on $\mathrm{CP^2}$.   The $U(1)$ rotation by $R_2$ is deformed to a  $U(1)_R'$ due to the twisting for both cases.  

The resulting 4 supersymmetric nonabelian  5d action on $\mathrm{R \times  CP^2} $ for the first case  is
\begin{align}
S_{\bf I }= \frac{k}{4\pi^2} \int_{\mathrm{R \times  CP^2}} d^5 x \sqrt{|g|}\ \tr \Big[ & -\frac14 F_{\m\n}F^{\m\n} +\frac{1}{2\sqrt{|g|}} \e^{\m\n\r\s\eta} J_{\m\n} \Big( A_\r \p_\s A_\eta -\frac{2i}{3}A_\r A_\s A_\eta \Big)
   \nn \\
 &  -\frac12 D_\m \phi_I D^\mu\phi_I +\frac14 [\phi_I,\phi_J]^2 -i\e_{abc}\phi_a[\phi_b,\phi_c]   -2  \phi_a^2
-\frac{13}{2} \phi_i^2 \nn \\
& -\frac{i}{2}\bar\l \g^\m D_\m \l -\frac{i}{2}  \bar\l \r_I[\phi_I,\l] -\frac18 \bar\l\g^{mn}\l J_{mn}+\frac34 \bar\l \r_{45}\l \Big] ,
\label{action1}
\end{align}
where $I=1,2,3,4,5$, $a=1,2,3$, $i=4,5$ and
\begin{align}
F_{\m\n} &= \p_\m A_\n -\p_\n A_\m -i[A_\m,A_\n], \nn \\
D_\m \phi_a &= \p_\m \phi_a - i [A_\m,\phi_a], \nn \\
D_\m \phi_i &= \p_\m \phi_i -i[A_\m,\phi_i]+3 V_\m \e_{ij}\phi_j ,\nn \\
D_\m \l &= \Big[\p_\m   +\frac14 \w_\m^{\r\s}\g_{\r\s} +\frac{3}{2}V_\m \r_{45} \Big]\l -i[A_\m,\l] .
\end{align}
Note that $J_{0m}=0$ and $J_{mn}$ is the K\"ahler 2-form on $\rm CP^2$. The  supersymmetric transformation is 
\begin{align} \d A_\m &= +i\bar\l \g_\m \e = -i\bar\e \g_\m \l , \nn \\
\d \phi_I &=-\bar\l \r_I \e = \bar\e \r_I \l ,\nn \\
\d \l  &= +\frac12 F_{\m\n}\g^{\m\n} \e + iD_\m \phi_I \r_I \g^\m \e -\frac{i}{2}[\phi_I,\phi_J]\r_{IJ}\e +   3i\e_{ij}\phi_i\r_j \e - 2\phi_I\r_I \tilde\e .  \end{align}
The supercharge  $Q$ is a singlet under $SU(3)$ isometry of $\mathrm{CP^2}$ and a doublet under $SU(2)_R$ with nontrivial $U(1)_R'$ charge. Thus the supergroup behind the first superconformal model would be $SU(1|2)$.

The 12 supersymmetric 5d action on $\rm R\times  CP^2$ for the second case is
\begin{align}
S_{\bf II} = \frac{k}{4\pi^2} \int_{\mathrm{R\times CP^2}} d^5 x\sqrt{|g|} \ \tr \Big[ & -\frac14 F_{\m\n}F^{\m\n} +\frac{1}{2\sqrt{|g|}} \e^{\m\n\r\s\eta} J_{\m\n} \Big( A_\r \p_\s A_\eta -\frac{2i}{3}A_\r A_\s A_\eta \Big)
   \nn \\
 &  -\frac12 D_\m \phi_I D^\mu\phi_I +\frac14 [\phi_I,\phi_J]^2 + \frac{i}{3}\e_{abc}\phi_a[\phi_b,\phi_c]   -2  \phi_a^2
-\frac{5}{2} \phi_i^2 \nn \\
& -\frac{i}{2}\bar\l \g^\m D_\m \l -\frac{i}{2}  \bar\l \r_I[\phi_I,\l] -\frac18 \bar\l\g^{mn}\l J_{mn} -\frac14 \bar\l \r_{45}\l \Big],
\end{align}
where $I=1,2,3,4,5$, $a=1,2,3$, $i=4,5$ and
\begin{align}
F_{\m\n} &= \p_\m A_\n -\p_\n A_\m -i[A_\m,A_\n], \nn \\
D_\m \phi_a &= \p_\m \phi_a - i [A_\m,\phi_a] ,\nn \\
D_\m \phi_i &= \p_\m \phi_i -i[A_\m,\phi_i]-V_\m \e_{ij}\phi_j ,\nn \\
D_\m \l &= \Big[\p_\m  +\frac14 \w_\m^{\r\s}\g_{\r\s} -\frac{1}{2}V_\m \r_{45} \Big]\l -i[A_\m,\l].
\end{align}
The supersymmetric transformation  is
\begin{align} \d A_\m &= i\bar\l \g_\m \e = -i\bar\e \g_\m \l ,\nn \\
\d \phi_I &=-\bar\l \r_I \e = \bar\e \r_I \l ,\nn \\
\d \l  &= +\frac12 F_{\m\n}\g^{\m\n} \e + iD_\m \phi_I \r_I \g^\m \e -\frac{i}{2}[\phi_I,\phi_J]\r_{IJ}\e + \e_{ij}\phi_i\r_j \e - 2\phi_I\r_I \tilde\e . \end{align}
The supercharge  $Q$ is a triplet under $SU(3)$ isometry of $\mathrm{CP^2}$ and a doublet under $SU(2)_R$ with nontrivial $U(1)_R'$ charge. Thus the supergroup behind the second superconformal model would be $SU(3|2)$.

\section{Properties of the 5d theories}

There are several properties of these 5d theories we like to focus in this section.   While we will not explore in detail in the present work,   instantons in our theories would play the Kaluza-Klein modes for the circle fiber as the case in the maximally supersymmetric 5d Yang-Mills theory on $R^5$.

The instanton number on $\mathrm{CP^2}$ is
\be \n = \frac{1}{8\pi^2} \int_{\mathrm{CP^2}} \Tr(F\wedge F)=\frac{1}{16\pi^2} \int_{\rm CP^2} d^4x \sqrt{|g|} \ \Tr F_{\m\n}\tilde F^{\m\n} . \ee
The eigenvalue  of the Hamiltonian of the 6d (2,0) theory on $\mathrm{R\times S^5}$ for an eigenstate  is  the conformal dimension of the corresponding operator.   Similarly the Hamiltonian for our 5d theories would have the conformal dimensions as the eigenvalues.
The abelian scalar field harmonics on $S^5$ is discussed in detail in later part of this section which shows that the lowest conformal dimension for the untwisted scalar field $\phi_a$ is two as expected. Upon $Z_k$ modding, first nontrivial Kaluza-Klein modes start with conformal dimension $k+2$. Such KK modes along the circle fiber is supposed to be represented by instantons in the 5d theory. As a single instanton has mass $4\pi^2/g^2_{YM}$ with our normalization $1/(4g_{YM}^2  )\Tr  F^2 $ where $F$ is $N\times N$ hermitian matrix valued two-form for   $U(N)$ gauge group and the KK modes has the additional mass $k$, the inverse coupling coefficient $1/g_{YM}^2$ is chosen to be $k/4\pi^2$. Instantons on $\mathrm{CP^2}$ have been explored  in Ref.~\cite{vafa}. It would be interesting to consider their work in our index calculation context.

While we do not have an argument for the quantization of the Chern-Simons term in the 3d sense, there is a simple argument in the 1d sense. Let us consider the abelian case with the spatial part of the vector potential $A=V$ so that  $F=dA=2J$, which has half instanton number and has $2\pi$ flux on non-contractiable two cycles of $\rm CP^2$. The Chern-Simons term in this background becomes
\be \frac{k}{4\pi^2}  \int_{\rm R\times CP^2}  d^5x   \ \frac12  \e^{\m\n\r\s\eta} J_{\m\n}\p_\r A_\s A_\eta \ \Rightarrow \int dt \ k A_0 \ee
The Chern-Simons level $k$ for this 1d   U(1) theory is again integer quantized as expected.

The quadratic Chern-Simons term has   been noted before\cite{ohlsson,kallen0}. A beautiful argument in Ref.~\cite{ohlsson} is that the $y$ independent field equation for the 3-form tensor field on $R\times S^5$ leads naturally to the presence of the quadratic Chern-Simons term.  Another argument is that the instantons are KK modes along the fiber direction and KK gauge field $V_p dx^p$ is a gauge field on $CP^2$ space with magnetic field $2J$. Whenever the instanton moves, it feels the background magnetic field and so the interaction term should be
proportional to $V_p dx^p/dt $ where $x^p$ is the position of a point like instanton on $CP^2$. The natural field theoretic expression is then the Chern-Simons term. It would be interesting to find more argument to support our choice of the coupling constant.
The full effect of this Chern-Simons term is not clear at this moment.

 The strongest coupling occurs for the $k=1$ case. For this value, there is no $Z_k$ modding and so the supersymmetry should be enhanced to the maximal value 32  with $OSp(8|2)$ supergroup.    
 For larger $k$, the story is more complicated. Let us examine the   6d Killing spinors after twisting but before dimensional reduction. 
Their dependence on the fiber direction $y$ would be depending on the $R_2$ charge. For the four supersymmetric case ({\bf I}),    $SU(3)$ singlet Killing spinors get split to  $\e_+\sim  e^{-it/2+0y},e^{it/2+3iy} $ and   $SU(3)$ triplet Killing spinors get split to $\e_+\sim e^{-\frac{i}{2}t -2i y}, e^{-\frac{i}{2}t +  iy } $ depending on $R_2$ charge. Upon the dimensional reduction, only $y$-independent modes and Killing spinors are realized explicitly.   For the $k=1$ case, all KK modes would be realized non-perturbatively and so are nontrivial Killing spinors.  For the $k=2$ case, the half of the triplet Killing spinors with even $y$ momentum  should be realized non-perturbatively in our 5d theory, and the total number of supersymmetries should be enhanced to  16. For the $k=3$ case, another half of the singlet Killing spinors can be  realized non-perturbatively instead, and so the total number of supersymmetries would get enhanced to 8. For the $k\ge 4$,  any  supersymmetry enhancement is not expected. For the 12 supersymmetric case ({\bf II}), 
$SU(3)$ triplet Killing spinors get split to    
$\e_+\sim e^{-\frac{i}{2}t+0y}, e^{-\frac{i}{2}t -iy}$ and  $SU(3)$ singlet Killing spinors get split to $\e_+\sim e^{-\frac{i}{2}t + 2iy}, e^{-\frac{i}{2}t+iy}$. For $k=2$ of this case, the half of singlet Killings spinors can be also realized and so the total supersymmetry would be 16. But there would be no more supersymmetry enhancement for $k\ge 3$.
   
One could ask whether instantons and anti-instantons without any other fields turned on are BPS in our 5d theories. Here we just consider   the selfdual or anti-selfdual gauge field strength, leaving the study of the instanton solution itself to the next paper. For the first case $({\bf I})$, we note that $\g_{1234}\e= -\e$ and only anti-instantons can be BPS. For the second case $({\bf II})$, both instantons and anti-instantons can be BPS. The amount of preserved super symmetries is interesting also. For the first case the anti-instantons preserve all of   4 susy. For the second case the anti-instantons preserve $4$ susy and instantons preserve $8$ susy.

While the instanton mass fixes the coupling constant, its 6d field theoretic origin can be read as follows: 
\begin{align}
 S_{6d} &= \int_{\mathrm{R\times S^5}} d^6x\sqrt{g} \ \Big( -\frac12  \p_\m  (\phi_I)_{6d}\p^\m (\phi_I)_{6d} + \cdots \Big) \nn \\
  \rightarrow S_{5d} &=  \frac{2 \pi r}{k}\int_{\mathrm{R\times CP^2}} d^5x \sqrt{g} \  \Big(-\frac12 \p_\m (\phi_I)_{6d} \p^\m (\phi_I)_{6d} + \cdots \Big) \nn \\ \rightarrow S_{5d} &=  \frac{k}{4\pi^2 r }\int_{\mathrm{R\times CP^2}} d^5x \sqrt{g} \  \Big(-\frac12 \p_\m (\phi_I)_{5d} \p^\m (\phi_I)_{5d} + \cdots \Big)  \end{align}
with the mass dimension for the 6d field $\phi_I$ being two and the mass dimension for the final 5d field $\phi_I$ being one.    The  scalar field is rescaled so that
\be (\phi_I)_{5d}=  \frac{2\pi\sqrt{2\pi}r}{k} (\phi_I)_{6d} . \ee

There is a Meyer's term in the potential which leads to nontrivial vacuum structure only for the first case. The classical supersymmetric vacua are given by the fuzzy 2-spheres which are given by the vacuum equation
\be ({\bf I}) \ \ -i[\phi_a,\phi_b] = 2\e_{abc}\phi_c. \ee
Naively, they would be D4 brane blown up to D6 brane whose world volume topology is $\rm R\times CP^2\times S^2$, and there would be corresponding giant graviton solutions. The exact nature of these vacua and their role will be explored in future.

The Gauss law in the $U(1)$ theory implies
\be \frac{k}{4\pi^2} D_m F^{m0} + \frac{k}{4\pi^2} e^{0mnpq}J_{mn}F_{pq}
  = 0.  \ee
Total charge should be zero in the compact $CP^2$. As $J$ is selfdual, the  anti-selfdual flux
\be F \sim e^1\wedge e^2- e^3\wedge e^4 \ee
seems possible without violation of the Gauss law but  it does not satisfy $dF=0$. The selfdual configuration $F=2J$ in the abelian theory with the instanton number one half   is not allowed due to the Gauss law without excitation of other fields. In nonabelian theories, there could be other charged matter fields and so the Gauss law could be satisfied nontrivially. There may be some monopole-like operator as in 3d case~\cite{seok0}.

While the second case has more supersymmetries, the first case is simpler as the Killing spinor is constant spinor on $\mathrm{CP^2}$. We will focus on the first case from now on. Still, it is hard to penetrate the  detail physics of the theory yet.  We do not see any restriction on the gauge group unlike the 6d theory~\cite{witten1}. 

Let us briefly mention the spectrum of the theory for first type (${\bf I}$) for abelian case. The detail spectrum for the scalar and fermion fields on $\mathrm S^5$ is given in~\cite{seok1}.   As the index calculation in the next section provides the detail of the spectrum on $\mathrm{CP^2}$, here we just focus on the scalar field spectrum. The spectrum of a scalar field of conformal dimension $2$ on $R\times S^5$ has the mass
\be 
(-\nabla_{S^5}^2 + 4) Y^{\ell_1,\ell_2} = (\ell_1+\ell_2+2)^2Y^{\ell_1,\ell_2}, \ -i \p_y Y^{\ell_1,\ell_2}= (\ell_1-\ell_2) Y^{\ell_1,\ell_2} . 
\ee
The highest weight of the given irreducible representation $Y^{\ell_1,\ell_2}$ would be $\ell_1 w_1 + \ell_2 w_2$ with two fundamental weights $w_1,w_2$ of $SU(3)$.
The dimension of the representation of the highest weight $(\ell_1,\ell_2)$ is $(\ell_1+1)(\ell_2+1)(\ell_1+\ell_2+2)/2$. 
For $\phi_{1,2,3}$ fields, there is no twisting. $Z_k$ modding puts the constraints $\ell_1-\ell_2=kn$ with integer $n$. The $y$-independent mode with $Y^{\ell,\ell}$ has the spectrum  on $CP^2$ as
\be (-\nabla_{\mathrm{CP^2}}^2 + 4)Y^{\ell,\ell}= 4(\ell+1)^2 Y^{\ell,\ell} , \ee
with the degeneracy $ (\ell+1)^3$. Note that the conformal dimension of this mode is $\varepsilon=2\ell+2$ and so it starts from 2 as we expect for the scalar field in the 6d theory. 
The first KK mode would with either $(\ell_1,\ell_2)=(k,0)$ or $(0,k)$. Both of them have the conformal dimension $\varepsilon=k+2$ with degeneracy $(k+1)(k+2)/2$. 

The twisted mode $\phi_4+i\phi_5$ has more complicated $y$-independent modes and KK modes. The $y$ independent mode is given by
$Y^{\ell,\ell+3}$ or $Y^{\ell+3,\ell}$ with conformal dimension $\varepsilon=2\ell+5$. One can do the similar analysis for the fermion field whose conformal dimension on $CP^2$ starts from $\varepsilon = 5/2$ as expected for the 6d fermion. The vector field analysis done in next section shows that its conformal dimension on $\rm CP^2$ starts from $\varepsilon=4$ not $3$ which is expected for the three-form tensor field in 6d. There may be no constant three form among the harmonics on $\rm S^5$. Instantons with perturbative effects should reproduce the KK modes. We hope to come back to these issues near future.
 
\section{Superconformal index}

We will now define the superconformal index on 6d (2,0) theory and analyze its properties upon the $Z_k$ modding introduced in section 2. Later we will relate this index with the 5d index on $\mathrm{R\times CP^2}$. The full 6d index would be obtainable from 5d computation involving the non-perturbative instanton states.
The superconformal index encodes the spectrum of BPS states of the radially quantized theory on $\mathrm{R\times S^5}$.
More precisely, the index we will define shortly counts the BPS states annihilated by a chosen supercharges $Q$ and its conjugate $S$ among 32 supercharges in 6d.
The chosen supercharge $Q$ satisfies the algebra
\be
    \{Q,S\} = \varepsilon -j_1 - j_2 -j_3 +2R_1 +2R_2 \equiv \Delta  ,
\ee
and hence the index will count BPS states saturating the bound $\Delta=0$.
Here the supercharge $Q$ has charges as $j_1=j_2=j_3=-\frac{1}{2}, R_1=R_2=-\frac{1}{2}$.

The superconformal index of the (2,0) theory is defined as
\be\label{index-definition}
    I(x,y_1,y_2,q) = {\rm tr}\Big[(-1)^F x^{\varepsilon+R_1}y_1^{j_1-j_2}y_2^{j_2-j_3}q^j\Big]   ,
\ee
where $x\!=\!e^{-\beta},y_1\!=\!e^{-i\gamma_1},y_2\!=\! e^{-i\gamma_2}$ denote the chemical potentials for the Cartan generators of the subalgebra commuting with $Q$, and $j=j_1+j_2+j_3-3R_2$.
This index for a single M5-brane and its gravity dual theory at large $N$ are studied in \cite{Bhattacharyya2}.
As the abelian (2,0) theory  is free, one can easily compute the index for the single M5-brane theory
by reading off BPS letters from the field content of the (2,0) theory.
The index of the $U(1)$ (2,0) theory is given by the Plethystic exponential of the single letter index $f$
\begin{align}
   &   I   = {\rm exp}\!\left[\sum_{n=1}^\infty \frac{1}{n}f(x^n,y_i^n,q^n)\right]  , \nn \\
     & \ \ \ \ \ \ \quad f(x,y_1,y_2,q) = \frac{x+x^2q^3-x^2q^2(1/y_1+y_1/y_2+y_2)+x^3q^3}{(1-xqy_1)(1-xqy_2/y_1)(1-xq/y_2)}   . 
\end{align}
The denominator comes from the derivatives, the first two terms of the numerator come from the scalar fields, three minus terms in the numerator come from the fermion fields, and the last term in the numerator comes from the fermion field equation. There is no contribution from the two-form tensor field.
One interesting limit of this index is to take $q\rightarrow 0$ limit where the index reduces to the half-BPS index
that is the index function of half-BPS states (preserving 16 supersymmetries).
In this limit, the letter index simply becomes $f=x$ and it reflects that only a single complex scalar $\phi_1-i\phi_2$  contributes to the index.
The $A_{N-1} $ non-abelian version of the half-BPS index \cite{Bhattacharyya1}   is already given by
\be\label{1/2-BPS-index}
    I_{{\rm 1/2-BPS}} = \prod_{m=1}^N\frac{1}{1-x^m}   .
\ee
This is the index we will reproduce in this section by calculating the perturbative part of the corresponding Euclidean path integral on $\mathrm{S^1\times CP^2}$.

Now we turn to the $Z_k$ modding of the superconformal index.
We introduced in section 2 the $Z_k$ quotient along the circular fiber direction $y$ twisted by $R_2$ rotation.
The $j$ corresponds to the rotation of this twisted $y$ direction.
The modding leaves only the $Z_k$ singlet states carrying $j=kn\ (n\in Z)$ charges and truncates all other states.
Accordingly, the index of the 6d theory with $Z_k$ quotient is defined as
\be\label{Zk-index}
    I_{Z_k} = {\rm tr}\Big[(-1)^F x^{\varepsilon+R_1}y_1^{j_1-j_2}y_2^{j_2-j_3}q^j\Big]\Big|_{j=kn} . 
\ee
When $k=1$, it reproduces the index for the (2,0) theory discussed above.
On the other hand, at infinite $k$ limit or zero coupling limit,
all the KK states with non-zero $j$ charge are truncated and
the index reduces to the 5d index counting the BPS states of the free theory on $\mathrm{R\times CP^2}$.
This limit is achieved by taking $q\rightarrow 0$ limit in the index computation.
Here, we note that this index at infinite $k$ coincides with the half-BPS index (\ref{1/2-BPS-index}) as two limits are achieved  identically by $q\rightarrow 0$.

We expect that the 5d index including the non-perturbative instanton states can reproduce the full 6d  superconformal index.
The 5d theory of the first case ({\bf I}) introduced in section 2 preserves the same supercharge $Q$ used to define the 6d index,
and, therefore, we can define the 5d index in the same way as the 6d index (\ref{index-definition}).
The perturbative states in 5d theory correspond to the $j$ singlet modes while the instanton states
realize the KK states with non-zero $j$ charge.
We thus identify the instanton number with the KK momentum number $j$.

The index can be considered as the Euclidean path integral of the 5d theory on $\mathrm{S^1\times CP^2}$
\be
    I(x,y_i,q) = \int_{\mathrm{S^1\times CP^2}} \mathcal{D}\Psi e^{-S_{\bf I}^E[\Psi]}   .
\ee
The twisted boundary condition along the time circle $S^1$ of radius $\beta r$ is considered.
The Euclidean version of the action (\ref{action1}) is given by
\be
    S_{\bf I}^E &=& \frac{k}{4\pi^2 r}\int_{\mathrm{S^1\times CP^2}} d^5x\sqrt{|g|}\ {\rm tr}\!\left[\frac{1}{4}F_{\mu\nu}F^{\mu\nu}+
     \frac{i}{2}\epsilon^{\mu\nu\lambda\rho\sigma}J_{\mu\nu}\left(A_\lambda\partial_\rho A_\sigma -\frac{2i}{3}A_\lambda A_\rho A_\sigma\right) \right.\nn \\
    && \ \ \ +\frac{1}{2}D_\mu \phi_ID^\mu \phi_I
    -\frac{1}{4}[\phi_I, \phi_J]^2+\frac{i}{r}\epsilon^{abc}[\phi_a,\phi_b]\phi_c + \frac{2}{r^2}(\phi_a)^2+\frac{13}{2r^2}(\phi_i)^2 \nn \\
    && \ \ \ \left. -\frac{i}{2}\lambda^\dagger\gamma^\mu D_\mu\lambda   - \frac{i}{2}\lambda^\dagger\rho_I[\lambda,\phi_I]-\frac{1}{8r}\lambda^\dagger J_{\mu\nu}\gamma^{\mu\nu}\lambda+\frac{3}{4r}\lambda^\dagger\rho_{45}\lambda \right] ,
\ee
where the fermion $\lambda$ is subject to the reality condition $\lambda = BC \lambda^*$ and the radius $r$ of $S^5$ is introduced again.
The twisted boundary condition shifts the time derivative such as
\be
    \partial_\tau \rightarrow \partial_\tau - \frac{\beta}{\beta r}R_1 - \frac{i\gamma_1}{\beta r}(j_1-j_2) -  \frac{i\gamma_2}{\beta r}(j_2-j_3) ,
\ee
and, from now on, we consider the time derivatives  as this shifted one.
The action is invariant under the supersymmetry transformation
\be
    \delta\phi_I &=& -\lambda^\dagger \rho_I\epsilon  ,\nn \\
    \delta A_\mu &=& -i\lambda^\dagger\gamma_\mu\epsilon ,\nn \\
    \delta\lambda &=& \frac{1}{2}F_{\mu\nu}\gamma^{\mu\nu}\epsilon -i D_\mu\phi_I\gamma^\mu\rho_I\epsilon -\frac{i}{2}[\phi_I,\phi_J]\rho_{IJ}\epsilon+ \frac{3}{r}\epsilon_{ij}\phi_i\rho_j\epsilon -\frac{2i}{r}\phi_I\rho_I\tilde\epsilon .
\ee
The supersymmetry parameter $\epsilon$ satisfies the conditions
\be
    D_\mu \epsilon = -\frac{i}{2r}J_{\mu\nu}\gamma^\nu\epsilon + \frac{1}{2r}\gamma_\mu\tilde \epsilon   , \quad
    \frac{3}{2}\rho^{45}\epsilon = - \frac{1}{4}J_{\mu\nu}\gamma^{\mu\nu}\epsilon +\frac{i}{2}\tilde\epsilon   , \quad \tilde \epsilon = i\rho^{45}\gamma_\tau\epsilon ,
\ee
and we found four solutions to these conditions,
\be
     \gamma_{12}\epsilon_+ = \gamma_{45}\epsilon_+ = -\rho^{45}\epsilon_+ = i\epsilon_+ , 
\ee
and its conjugation $\epsilon_- = BC \epsilon_+^*$. It turns out that the four Killing spinors are convariantly constant on $\mathrm{CP^2}$
\be
    D_m\epsilon_\pm = 0     \ (m=1,2,3,4) .
\ee

We would like to evaluate the superconformal index using the localization technique.
The localization would lead  to the path integral over the instanton configuration on $\mathrm{CP^2}$ base. 
We leave the calculation of the nonperturbative instanton contributions for future work.

At infinite $k$, the gaussian integral of the quadratic equations produces the exact result.
For convenience, let us divide the field content to a vector multiplet and an adjoint hypermultiplet
(though there is no notion of the hypermultiplet as the theory preserves only 4 supercharges).
We first pick up a complex supercharge $Q$ corresponding to $\rho_{12}\epsilon = -i\epsilon$ and decompose the spinors as
\be
    \epsilon = \left(\begin{array}{c}\!\!\epsilon_-\!\! \\ \!\!\epsilon_+\!\!\end{array}\right)\otimes\left(\begin{array}{c}\!\!1\!\!\\\!\!0\!\!\end{array}\right)
      , \quad \lambda = \left(\begin{array}{c}\!\! \chi^1  \!\! \\ \!\! \chi^2 \!\! \end{array}\right) \otimes \left(\begin{array}{c}\!\!1\!\!\\\!\!0\!\!\end{array}\right)
    +\left(\begin{array}{c}\!\! \psi^1  \!\! \\ \!\! \psi^2 \!\! \end{array}\right) \otimes \left(\begin{array}{c}\!\!0\!\!\\\!\!1\!\!\end{array}\right)   .
\ee
Then the vector multiplet consists of $A_\mu,\chi,\phi_3$ and the hypetmultiplet consists of two complex scalar $q^A$ and a complex fermion $\psi$
defined as
\be
    q_1 \equiv \frac{1}{\sqrt{2}}(\phi_4-i\phi_5)   , \quad q_2 \equiv \frac{1}{\sqrt{2}}(\phi_1+i\phi_2)   , \quad \psi \equiv \psi^2   .
\ee
The action with the new fields becomes
\be
   S_{\bf I}^E &=& \frac{k}{4\pi^2}\int_{\mathrm{R\times CP^2}} d^5x\sqrt{|g|}\ {\rm tr}\!\left[\frac{1}{4}F_{\mu\nu}F^{\mu\nu}+
     \frac{i}{2}\epsilon^{\mu\nu\lambda\rho\sigma}J_{\mu\nu}\left(A_\lambda\partial_\rho A_\sigma -\frac{2i}{3}A_\lambda A_\rho A_\sigma\right) \right.\nn \\
     &&+\frac{1}{2}D_\mu \phi_3D^\mu \phi_3+|D_\mu q^A|^2
    +\frac{2}{r^2}(\phi_3)^2+\frac{4}{r^2}|q^2|^2 + \frac{13}{r^2}|q^1|^2
    \nn \\
    &&+|[\phi_3,q^A]|^2+\frac{1}{2}|[q^A,\bar{q}_A]|^2+\frac{1}{2}(\sigma^I)^A_{\ \ B}(\sigma^I)^C_{\ \ D}[q^B,\bar{q}_A][q^D,\bar{q}_C] -\frac{6}{r}\phi_3[q^2,\bar{q}_2]  \nn \\
    &&-\frac{i}{2}\chi^\dagger\gamma^\mu D_\mu\chi -i\psi\gamma^\mu D_\mu\psi -\frac{1}{8r}\chi^\dagger J_{\mu\nu}\gamma^{\mu\nu}\chi -\frac{1}{4r}\psi^\dagger J_{\mu\nu}\gamma^{\mu\nu}\psi+\frac{3i}{4r}\chi^\dagger\sigma^3\chi+\frac{3i}{2r}\psi^\dagger\psi \nn \\
    && \left.- \frac{i}{2}\chi^\dagger[\phi_3,\chi]+i\psi^\dagger[\phi_3,\psi] +\sqrt{2}i\psi^\dagger[\chi_A,q^A] -\sqrt{2}i[\bar{q}_A,\chi^\dagger]\psi\right]   ,
\ee
where $\sigma^{I=1,2,3}$ are the Pauli matrices.

Before performing the path integral, let us first fix the gauge following \cite{kinney}.
We choose the Coulomb gauge $D^m A_m =0$ and impose the residual gauge fixing condition as $\frac{d}{d\tau}\alpha=0$
where $\alpha \equiv \frac{1}{\omega_{CP^2}} \int_{CP^2}A_\tau$ is the s-wave component (or holonomy) of $A_\tau$.
The holonomy $\alpha$ is the only zero mode of the quadratic action.
The residual gauge fixing introduces the Haar measure to the path integral.
Thus the index at large $k$ becomes the integral of the 1-loop determinant by the holonomy $\alpha$
\be
    I  = \frac{1}{N!}\int \prod_{i=1}^N[\frac{d\alpha_i}{2\pi}]\prod_{i<j}^N\left[2\sin\left(\frac{\alpha_i-\alpha_j}{2}\right)\right]^2\times I_{1-loop}   .
\ee

To obtain the 1-loop determinant, we will use the various $\mathrm{CP^2}$ harmonics carrying electric charges $R_2$.
Some of them are constructed in \cite{Pope,seok1}.
Let us first focus on the scalars in the hypermultiplet.
The scalars have the following quadratic terms
\be
    \bar{q}_1\left[-D^2_\tau - D^mD_m +\frac{13}{r^2}\right]q^1+\bar{q}_2\left[-D^2_\tau - D^mD_m +\frac{4}{r^2}\right]q^2   ,
\ee
where the time derivative is
\be
    D_\tau = \partial_\tau -i[\alpha,\ \ ] - \frac{\beta}{\beta r}R_1 - \frac{i\gamma_1}{\beta r}(j_1-j_2) -  \frac{i\gamma_2}{\beta r}(j_2-j_3)   .
\ee
We need to use the charged $SU(3)$ harmonics $Y^{l+3R_2,l}$ if $R_2>0$ or $Y^{l,l+3|R_2|}$ if $R_2<0$ according to $R_2$ charges of the scalar fields.
Here, the charged harmonics $Y^{l_1,l_2}$ carries $R_2$ charge $\frac{l_1-l_2}{3}$.
Then the corresponding harmonics are $Y^{l,l+3}$ for $q^1$ and $Y^{l,l}$ for $q^2$ respectively, and they diagonalize the quadratic equation.
The 1-loop determinant of the hyper scalars becomes
\be
   {\rm det}_{H,b} \!\!&\!\!=\!\!&\!\!\prod_{\alpha\in root}\prod_{l=0}^\infty\prod_{m_1,m_2\in (l,l+3)}\sin\left(\frac{\alpha +m_i\gamma_i+i(2l+5)\beta}{2}\right)\sin\left(\frac{\alpha +m_i\gamma_i-i(2l+5)\beta}{2}\right) \nn \\
    && \!\!\! \!\!\!\!\!\!\!\!\! \times\!\! \prod_{\alpha\in root}\prod_{l=0}^\infty\prod_{m_1,m_2\in (l,l)}\sin\left(\frac{\alpha +m_i\gamma_i+i(2l+1)\beta}{2}\right)\sin\left(\frac{\alpha +m_i\gamma_i-i(2l+3)\beta}{2}\right)
     . \qquad
\ee
where $m_i\gamma_i = m_1\gamma_1+m_2\gamma_2$ and $m_i$ denote the two Cartan charges of $(l_1,l_2)$ representation of $SU(3)$ isometry.

For the complex fermion $\psi$, we introduce the four spinor basis on $CP^2$
\be\label{spinor-basis}
    \Psi_1 = Y^{l,l+3}\epsilon_+ \,  , \quad \Psi_2 =\gamma^\tau \gamma^m D_m Y^{l,l+3} \epsilon_+ \, , \quad \Psi_3 = Y^{l,l}\epsilon_- \,  , \quad \Psi_4 = \gamma^\tau\gamma^mD_m Y^{l,l}\epsilon_-   , 
\ee
where $Y^{l_1,l_2}$ is the charged $SU(3)$ harmonics defined above.
These four basis can diagonalize the fermion quadratic action
\be
    \psi^\dagger\left[-i\gamma^\tau D_\tau  -i  D^m\gamma_m  -\frac{1}{4r} J_{mn}\gamma^{mn} + \frac{3i}{2r}\right]\psi\, .
\ee
One then obtains the 1-loop determinant for the fermion field in the hypermultiplet
\be
   \hspace{-1cm} &&{\rm det}_{H,f} =\prod_{\alpha\in root}\prod_{l=0}^\infty\prod_{m_1,m_2\in (l,l+3)}\sin\left(\frac{\alpha +m_i\gamma_i+i(2l+5)\beta}{2}\right)\sin\left(\frac{\alpha +m_i\gamma_i-i(2l+5)\beta}{2}\right) \nn \\
   \hspace{-1cm} &&\!\!\!\!\times\!  \sin\left(\!\frac{\alpha-3i\beta}{2}\!\right)\!\!\! \prod_{\alpha\in root}\!  \prod_{l=1}^\infty\prod_{m_1,m_2\in (l,l)}\!\!\!\!\!\!\! \sin\!\left(\!\frac{\alpha +m_i\gamma_i+i(2l+1)\beta}{2}\!\right)\sin\!\left(\!\frac{\alpha +m_i\gamma_i-i(2l+3)\beta}{2}\!\right)
   . \qquad
\ee
The first line corresponds to the 1-loop determinant from $\Psi_1,\Psi_2$ and the second line is from $\Psi_3,\Psi_4$.
Combining the complex scalar and the fermion contributions, the final 1-loop determinant of the hypermultiplet is given by
\be
    \frac{{\rm det}_{H,f}}{{\rm det}_{H,b}} = \prod_{\alpha\in root}\frac{1}{\sin\left(\!\frac{\alpha+i\beta}{2}\!\right)} 
    =x^{\varepsilon_0} {\rm exp}\!\left[\sum_{n=1}^\infty\sum_{i,j}^N\frac{1}{n}x^n e^{ni\alpha_{ij}}\right] .
\ee
where $\varepsilon_0= \frac{1}{2}N^2$ is the Casimir energy for the hypermultiplet.

Let us move on to the vectormultiplet contribution.
It is straightforward to compute the fermionic contribution by using the same spinor basis (\ref{spinor-basis}).
The quadratic equation for $\chi^1$ is given by
\be
    (\chi^1)^\dagger \left[-i\gamma^\tau D_\tau -i\gamma^m D_m -\frac{1}{4r}J_{mn}\gamma^{mn}+\frac{3i}{2r}\right]\chi^1 .
\ee
The corresponding 1-loop determinant becomes
\be
    \hspace{-1cm} &&{\rm det}_{V,f} =\prod_{\alpha\in root}\prod_{l=0}^\infty\prod_{m_1,m_2\in (l,l+3)}\sin\left(\frac{\alpha +m_i\gamma_i+i(2l+6)\beta}{2}\right)\sin\left(\frac{\alpha +m_i\gamma_i-i(2l+4)\beta}{2}\right) \nn \\
   \hspace{-1cm} &&\!\!\times\! \sin\!\!\left(\!\frac{\alpha-2i\beta}{2}\!\right)\!\!\!\prod_{\alpha\in root}\prod_{l=1}^\infty\prod_{m_1,m_2\in (l,l)}\!\!\!\!\!\!\sin\!\!\left(\!\frac{\alpha +m_i\gamma_i+i(2l+2)\beta}{2}\!\right) \!\sin\!\!\left(\!\frac{\alpha +m_i\gamma_i-i(2l+2)\beta}{2}\!\right)
     . \qquad
\ee
The first line is again the 1-loop determinant of $\Psi_1,\Psi_2$ and the second line is from $\Psi_3,\Psi_4$.

The quadratic action of the vector field is
\be\label{vector-action}
     && \frac{1}{2}F_{\mu\nu}F^{\mu\nu}+i\epsilon^{\mu\nu\lambda\rho\sigma}A_\mu\partial_\nu A_\lambda J_{\rho\sigma} = (D_m A_\tau)^2  + 2A_\tau \partial_\tau D_m A^m\nn \\
     & & \ \ \ \ -A_m\!\left(D_\tau^2 \delta^m_n + D^2\delta^m_n -D^mD_n -6\right)\! A^n +4iA_\tau D_m A_n J^{mn} -2i A_m D_\tau A_n J^{mn}. 
   \qquad
\ee
We find that the following vector harmonics form the complete basis of the 5 vector components
\be
    \mathcal{A}_\tau = Y^{l,l}  ,   \quad \mathcal{A}^1_m = D_m Y^{l,l} \ , \quad \mathcal{A}^2_m = J_{mn}D^n Y^{l,l}   , \quad \mathcal{A}^3_m = \epsilon^\dagger_- \gamma_m \gamma^n D_n Y^{l,l+3}\epsilon_+.
\ee
Here, $\mathcal{A}_\tau, \mathcal{A}^1, \mathcal{A}^2$ are real vectors and $\mathcal{A}^3$ is a complex vector.
As we have already taken into account the zero mode of $A_\tau$, which gives the holonomy $\alpha$ and Haar measure of the gauge group,
the range of the harmonics $Y^{l,l}$ is therefore $l>0$.
Under the Coulomb gauge $D^m A_m=0$, we can turn off the modes corresponding to $\mathcal{A}^1_m$.
The other two real vectors $\mathcal{A}_\tau,\mathcal{A}^2_m$ mix each other in the quadratic action.
Taking into account the determinant factors from the gauge fixing procedure, we obtain the 1-loop determinant for the real vectors
\be
    \prod_{\alpha\in root}\prod_{l=1}^\infty\prod_{m_1,m_2\in (l,l)}\left[\sin\left(\frac{\alpha +m_i\gamma_i+i(2l+2)\beta}{2}\right)\sin\left(\frac{\alpha +m_i\gamma_i-i(2l+2)\beta}{2}\right)\right]^{\frac12}\!\! . 
\ee
The complex vector $\mathcal{A}^3$ is an eigenvector of the quadratic equation (\ref{vector-action})
and its 1-loop determinant is
\be
    \prod_{\alpha\in root}\prod_{l=0}^\infty\prod_{m_1,m_2\in (l,l+3)}\sin\left(\frac{\alpha +m_i\gamma_i+i(2l+6)\beta}{2}\right)\sin\left(\frac{\alpha +m_i\gamma_i-i(2l+4)\beta}{2}\right) .
\ee
We then collect the fermion and the vector contributions as well as the contribution from a scalar field $\phi^3$.
After the huge cancellation between the fermionic and bosonic contributions, we finally find that the 1-loop determinant of the vector multiplet is trivial
\be
    \frac{{\rm det}_{V,f}}{{\rm det}_{V,b}} = 1 .
\ee

Combining the contributions from the vector and the hypermultiplet, we obtain the following superconformal index at infinite $k$
\be
    I(x,y_1,y_2)_{k\rightarrow \infty} &=& \frac{x^{\varepsilon_0}}{N!}\int \prod_{i=1}^N[\frac{d\alpha_i}{2\pi}]\prod_{i<j}^N\left[2\sin\left(\frac{\alpha_i-\alpha_j}{2}\right)\right]^2{\rm exp}\left[\sum_{n=1}^\infty\sum_{i,j}\frac{1}{n}x^n e^{ni\alpha_{ij}}\right] \nn \\
    &=&x^{\varepsilon_0}\prod_{m=1}^N\frac{1}{1-x^m} .
\ee
It follows that the index receives the contributions from the states formed by a single letter $\phi_1-i\phi_2$.
This result agrees with the 6d superconformal index at infinite $k$ and, therefore, agrees with the half-BPS index (\ref{1/2-BPS-index}).
We believe that the full superconformal index at finite $k$ can be calculated by including the instanton contribution.

\section{Supergravity}

Let us briefly consider the $\mathrm{AdS_7\times S^4}$ geometry corresponding to the 6d (2,0) theory~\cite{adscft}. In case we need the complete $\mathrm{AdS_7}$ geometry with $S^5$ boundary. The maximally supersymmetric $\mathrm{AdS_7\times S^4}$ geometry is
\begin{align} & ds^2  = R^2 (-\cosh^2\r dt^2 +d\r^2 +\sinh^2\r ds^2_{S^5} )+ \frac14 R^2  ds^2_{S^4}  , \nn \\
& F_4  \sim N \e_4 , \   
 R/\ell_p   = 2(\pi N)^{1/3} .
 \end{align}
The 5d unit sphere and 4d unit sphere are modded by
\be \frac{\mathrm{S^5\times S^4}}{Z_k}.  \ee
The metrics on $S^5$ and $S^4$ are, respectively, 
\begin{align} ds_{S^5}^2 &= ds_{CP^2}^2 + (dy'+ V)^2 ,\nn\\
ds_{S^4}^2 &= d\vartheta^2 + \sin^2 \vartheta d\chi'^2 +\cos^2 \vartheta ds_{S^2} .
\end{align}
where $\chi'$ is the phase corresponding to the phase of   $\phi_4+i\phi_5$ and $dV= 2J$ is the K\"ahler 2-form on $\mathrm{CP^2}$.
The $Z_k$ modding for the first and second cases  are
\begin{align} ({\bf I}) \ & (y',\chi') \sim (y',\chi')+ \frac{2\pi}{k} (1,3) , \nn \\
({\bf II})  \ &  (y',\chi') \sim (y',\chi')+ \frac{2\pi}{k} (1,-1) . \end{align}
Let us focus on the first case with the change of coordinates  to
\be y'= \frac{y}{k}, \ \chi'= \chi + \frac{3y}{k},  \ee
with $y\in [0,2\pi]$ and $\chi\in [0,2\pi]$.
The geometry becomes
\begin{align} \label{moddedg} ds^2 &=R^2\Big[-\cosh^2\r dt^2 + d\r^2 +\sinh^2\r  ds_{CP^2}^2 + \frac{1}{k^2} \sinh^2\r (dy + kV)^2 ] \Big] \nn \\
& \ \ \ + \frac{R^2}{4}\Big[  d\vartheta^2 + \sin^2\vartheta ( d\chi + \frac{3dy}{k})^2   + \cos^2\vartheta  ds^2_{S^2} \Big], 
  \nn \\
F &\sim N ( {\cal V}_{S^4}+  \frac{3}{k} \sin\vartheta \cos^2\vartheta \
d\vartheta\wedge  dy \wedge {\cal V}_{S^2}) ,
\end{align}
where
\begin{align} 
{\cal V}_{S^4} &= \sin\vartheta \cos^2\vartheta  \  d\vartheta \wedge d\chi \wedge {\cal V}_{S^2}.  
\end{align}
where ${\cal V}_{S^2} $ is the volume form of a unit  2-sphere.

The corresponding Type IIA geometry  can be obtained by the relation:
\begin{align} \label{metrel} ds_{11}^2 &= e^{-2\s/3  } ds_{10}^2 + e^{4\s/3}(dy +{\cal A})^2,  \nn \\
F^4_{11} &= e^{4\s/3} F^4_{10}+  e^{\s/3} F^3_{10}\wedge dy  .\end{align}
Some of NS-NS fields of $\s,g_{MN}, B_{MN} $ and  R-R fields $C_\m, C_{\m\n\r}$ are nonvanishing as  $C_Mdx^M= {\cal A}$ and $e^{\s/3}F^3_{10}=e^{\s/3}dB=  \frac{3N}{k} \sin\vartheta \cos^2\vartheta
d\vartheta\wedge {\cal V}_{S^2} $. 
The metric (\ref{moddedg}) containing $(dy +kV)^2$ and $(d\chi +3dy/k)^2$ becomes
\begin{align}
\frac{R^2}{4k^2}(  4\sinh^2\r +9\sin^2\vartheta ) (dy +   {\cal A} )^2 + \frac{ R^2\sinh^2\r\sin^2\vartheta}{4\sinh^2\r+9\sin^2\vartheta} (d\chi- 3V)^2 , \end{align}
where
\be {\cal A}  = k \frac{ 4 \sinh^2\r\  V + 3\sin^2\vartheta d\chi}{4\sinh^2\r + 9\sin^2\vartheta} .   \ee
Thus the relation (\ref{metrel}) implies
\be e^{4\s/3} = \frac{R^2}{4k^2}( 4\sinh^2\r + 9\sin^2\vartheta) , \ee
and
\begin{align} e^{-2\s/3} ds^2_{10} = & +R^2   [ -\cosh^2\r dt^2 + d\r^2 + \sinh^2\r d s^2_{CP^2}] \nn \\
& + \frac{R^2}{4} (d\vartheta^2 + \cos^2\vartheta ds^2_{S^2}) +\frac{R^2\sinh^2\r\sin^2\vartheta}{4\sinh^2\r +9\sin^2\vartheta} (d\chi-3V)^2. \end{align}
The field strength are
\begin{align}  F^4_{10} & = N e^{-4\s/3} {\cal V}_S^4, \nn \\
F^3_{10} & =\frac{3N}{k} e^{-\s/3}  \sin\vartheta d\vartheta \wedge {\cal V}_{S^2} .
\end{align}
Note that $F^4_{10}$ is for the D4 branes and $F^3_{10}$ is for the D6 branes.

The radius of the circle fiber $y$ is of order
\be   e^{4\s/3} \sim \frac{  N^{1/3}}{k}     \sinh\r . \ee
As we divide the $AdS_7$ space, we do not have a small compact circle and
so it is hard to say the theory has been reduced to the type IIA theory.
However the above radius says  that the M-theory description is valid for
$1\le k \lesssim N^{1/3}$.  
Since the dilation field diverges at the boundary, the ultraviolet physics at the boundary is the 6d physics. 
The string frame metric (\ref{metrel}) in type IIA gives
\begin{align}
 ds^2_{10} &= \frac{R^3}{2k}\Big[  (-\cosh^2\r dt^2 + d\r^2 + \sinh^2\r d s^2_{CP^2}) \nn \\
 &  \ \ \ +
  \frac14 (d\vartheta^2 + \cos^2\vartheta ds^2_{S^2})+  \frac{\sinh^2\r\sin^2\vartheta}{4\sinh^2\r +9\sin^2\vartheta} (d\chi-3V)^2 \Big] .
  \end{align}
The curvature scale of the type IIA theory is of order $\sqrt{R^3/2k} \sim
\sqrt{N/k}$ which is large  when 't Hooft coupling $\l=N/k$ is large.

\section{Conclusion and Discussion}

We have found the supersymmetric Yang-Mills Chern-Simons theories on $\rm R\times CP^2$ which arise  from  the $Z_k$ modding of the 6d$(2,0)$ theory on $\rm R\times S^5$   with additional twistings along the $R$ symmetry direction. Depending on the twisting, the number of supersymmetries  can be   4 or 12. Here for simplicity we have focused the analysis for 4 supersymmetric case with the supersymmetric spinor parameter which is a singlet under the $SU(3)$ isometry of $\rm CP^2$. 
The fluctuation analysis shows that the fields have the right conformal dimension as expected from the 6d consideration. Supergravity analysis shows that there are M-theory region and type IIA region and weakly coupled region even though the boundary between first two regions is not that distinct.  

We have argued that the number of   supersymmetries get enhanced for $k=1,2,3$ cases when the nonperturbative effects are included. 
As there is a discrete coupling constant, there is a good chance that our theories are finite in UV and represent the 6d theory completely once the nonperturbative effects are included.

Our theories are good stepping stones for calculating    the index function of the 6d (2,0) theory and we hope to report the result in near future. There seems to be several interesting ideas to pursue from the current point. There may be many BPS objects in the theory which is not apparent in
first glance. The $N^3$ degrees of freedom on the 6d (2,0) theory~\cite{klebanov} have been studied from various points of view~\cite{seok1,kallen1,bolognesi} and our theory may provide a further evidence.

\section*{Acknowledgment}

We are very grateful to Dongmin Gang, Seok Kim, Sung-Soo Kim, Eunkyung Koh, Sunil Mukhi, and Jaemo Park     for discussions.   K.L. thanks    Newton Institute for Mathematical Science  and Aspen Center for Physics for hospitality where the part of this work is done. This work  is supported by the National Research Foundation of Korea  Grants No. 2010-0007512 (HK), 2006-0093850 (KL), 2009-0084601 (KL), and 2005-0049409 (KL) through the Center for Quantum Space-Time(CQUeST) of Sogang University.

\appendix

\section{convention for metrics and gamma matrices}

The space-time metric has the mostly positive signature.
The metric tensors on $\mathrm{CP^2}$ and $\mathrm{S^5}$ are, respectively,
\begin{align} ds^2_{\mathrm{CP}^2} & = d\r^2 +\frac{\t_3^2}{4}\sin^2\r \cos^2\r \   + \frac{ \t_1^2+\t_2^2 }{4} \sin^2\r    , \nn \\
ds^2_{\mathrm{S}^5} & = ds^2_{\mathrm{CP}^2}+ (dy+V)^2,
  \ \   \ V= \frac{\t_3}{2} \sin^2\r,
\end{align}
where $y$ is the $U(1)$ fiber direction. The left-invariant $SU(2)$ 1-forms are
\begin{align}
\t_1 &= -\sin\psi d\theta +\cos\psi\sin\theta d\varphi, \nn \\
\t_2 &= +\cos\psi d\theta+\sin\psi\sin\theta d\varphi, \nn \\
\t_3 &= + d\psi+\cos\theta d\varphi,
\end{align}
such that $ d \t_i = \frac12 \e_{ijk} \t_j\wedge \t_k $.
  The range of variables are
  $\r\in[0,\frac{\pi}{2}], \theta\in[0,\pi], \varphi\in[0,2\pi],\psi\in[0,4\pi]$  and $y\in[0,2\pi]$. The volumes of $\mathrm{CP}^2$  and $\mathrm{S^5}$ are  $\pi^2/2$ and   $\pi^3$, respectively.

The vierbein $e^m=e^m_p dx^p$ for $\mathrm{CP}^2$ is  %
\begin{align}
& e^1= d\r, \ e^2= \frac{\t_3}{2}   \sin\r \cos\r  , \ e^3= \frac{\t_1}{2} \sin\r    , \ e^4=\frac{\t_2}{2} \sin\r.
\end{align}
Their inverse  $e_m = e_m^p \p_p$ is
\be  e_1= \p_\r, \ e_2 =\frac{2\tilde\t_3}{\sin\r\cos\r} , \ e_3= \frac{2\tilde\t_1}{\sin\r}, \ e_4=\frac{2\tilde\t_2}{\sin\r},
\ee
where
\begin{align} \tilde \t_1 &= -\sin\psi\p_\theta +\frac{\cos\psi}{\sin\theta}(\p_\varphi
-\cos\theta\p_\psi), \nn \\
 \tilde \t_2 &= +\cos\psi\p_\theta +\frac{\sin\psi}{\sin\theta}(\p_\varphi
-\cos\theta\p_\psi) ,\nn \\
\tilde \t_3 &= + \p_\psi .
\end{align}
The K\"ahler 2-form on $\mathrm{CP}^2$ is
\be J = \frac12 J_{mn}e^m\wedge e^n=\frac12 dV = e^1\wedge e^2 + e^3\wedge e^4. \ee
The spin-connection for the $\mathrm{CP}^2$ is
\begin{align}
& w^{12} = -\frac{\t_3}{2}\cos 2\r , \ w^{34} = +\frac{\t_3}{2}( 1+\sin^2\r),\nn \\
& w^{23} = w^{41}=  +\frac{\t_2}{2}\cos\r, \ w^{31}= w^{42} = +\frac{\t_1}{2}\cos\r.
\end{align}
The vierbein on $S^5$ is
\begin{align}
&  E^m = e^m \  (m=1,2,3,4), \ \   E^5=
dy + V_p dx^p. \end{align}
The inverse vierbein on   $\mathrm{S}^5$   is
\be
  E_m = e_m - e_m^p V_p \p_y  \ (m=1,2,3,4) , \   \ E_5= \p_y.
\ee
The spin connection for $\mathrm{S}^5$ is
\be
  W^{mn} = w^{mn} - J^{mn} E^5,  \   W^5_{\ m} = J_{mn} e^n.
\ee

Our notation for the Minkowski space-time gamma matrices for 6d and 5d is  as follows:
\begin{align}  (5d)\  &\  \g^0=1_2\otimes i\s_2, \ \g^{1,2,3}=\s_{1,2,3}\otimes \s_1, \ \g^4=1_2\otimes \s_3, \ \g^{01234}=i1_4,  \nn \\
(6d)\ & \ \G^\m= \g^\m\otimes \s_1 \ (\m=0,1,\cdots 4), \ \G^5=1_4\otimes \s_2, \ \G^7=\G^{01\cdots 5} = -1_4\otimes \s_3 .
\end{align}
The 6d spinor field $\l$ and the supersymmetric parameter $\e$ have
the opposite chirality so that $\G^7\l=\l, \G^7\e=-\e$.
With $B=i\s_2\otimes \s_1$, we get $B\g^\m B^{-1}= -\g^{\m^*}=-\g_\m^T$.
The spinors transform as ${\bf 4}$ of $Sp(2)_R=SO(5)$ symmetry and the 5d Euclidean gamma matrices on ${\bf 4}$ are
\be \r_{1,2,3}= \s_{1,2,3}\otimes \s_3,\ \r_4=1_2\otimes \s_2,\  \r_5= 1_2\otimes \s_1 .
\ee
Our choice of Cartan for $Sp(2)_R$ is $R_2 \sim \frac12\r_{45}$ and $R_1   \sim \frac12 \r_{12}$ to fermionic fields.
The charge conjugation operator acting on ${\bf 4}$ of $Sp(2)_R$ is  $C=i\s_2\otimes \s_1$ such that $C\r_I C^{-1}= \r_I^T$.
With $\hat B = B\otimes \s_3$, we get $\hat B \G^M \hat B^{-1} = \G^{M*} = \G_M^T$.
We require  the the reality conditions on the spinors to be
\be \l = -\hat B C \l^*, \ \e=\hat B C     \e^* \ \ \Longrightarrow \ \ \l = BC\l^*, \ \e= BC\e^* \ee
on four component spinors.

\section{Killing spinors}

The Killing spinors~\cite{seok1,Pope} on $R\times S^5$ are defined as follows:
\begin{align}
 \hat\nabla_M \e_\pm  = \frac{i}{2} \G_M \tilde \e_\pm = \pm \frac{i}{2} \G_M   \G_0\e_\pm,
 \end{align}
and $\e_\pm = \hat B C\e^*_\mp $ and $\tilde \e_\pm = \pm \G_0\e$.
Here we will be loose about 8 and 4 component spinors as the chirality condition $\G^7\l=\l, \G^7\e=-\e, \G^7\tilde\e=\tilde\e$ leaves no ambiguity. The covariant derivative  to the spinor on $\mathrm S^5$ given as
\be \nabla_M\e = (\p_M + \frac14 W^{AB}_M \G_{AB}) \e .\ee

The covariant derivative on spinors in  $\mathrm{S}^5$ can be expressed in terms of that on $\mathrm{CP}^2$ plus the derivative along the circle fiber.
\begin{align}
  \hat \nabla_0\e &\equiv \p_t \e = \frac{i}{2} \g_0\tilde \e , \nn \\
  \hat \nabla_m\e & \equiv\Big[ \nabla_m -V_m \p_y +\frac12 J_{mn}\G^{n5} \Big]\e  \nn = \Big[\nabla_m -V_m\p_y +\frac{i}{2}J_{mn}\g^n\Big]\e =\frac{i}{2}\g_m\tilde\e, \\
 \hat\nabla_5\e & \equiv \Big[\p_y -\frac14 J_{mn}\G^{mn}\Big]\e = \Big[\p_y -\frac14 J_{mn} \g^{mn}\Big] \e=\frac12 \tilde\e, 
\end{align}
where $m=1,2,3,4$ and $V=V_m e^m = V_\m dx^\m, J = \frac12 J_{mn}e^m\wedge e^n $ and $\nabla_m=e_m^\m \nabla_\m$ is the spinor covariant  derivative   on $\mathrm{CP}^2$. The Killing spinor equation is solved with
\be \tilde \e = \pm \G_0\e = \pm \g_0\e  \ee
The covariant derivative on the gaugino field is
\begin{align}
  \hat \nabla_0\l &\equiv \p_t \l ,\nn \\
  \hat \nabla_m\l & \equiv\Big[ \nabla_m -V_m \p_y +\frac12 J_{mn}\G^{n5} \Big]\l  \nn = \Big[\nabla_m -V_m\p_y -\frac{i}{2}J_{mn}\g^n\Big]\l  , \\
 \hat\nabla_5\l & \equiv \Big[\p_y -\frac14 J_{mn}\G^{mn}\Big]\l = \Big[\p_y -\frac14 J_{mn} \g^{mn}\Big]  \l .
\end{align}

Let us split the spinors to eigenspinors $\g_{12}\e^{s_1s_2}=is_1\e^{s_1s_1}, \g_{34}\e^{s_1\s_2}=is_2\e^{s_1\s_2}$. Note that $\g^0\e^{s_1s_1}= i\g^{1234}\e^{s_1s_2}= -is_1s_2\e^{s_1s_2}$. One solution of the Killing spinor is
\be  ({\bf I})\ \ \ \e_+\sim e^{-\frac{i}{2}t + \frac{3i}{2}y}\e_0^{++} \ee
with a constant spinor $\e_0^{++}$. It is singlet under the $SU(3)$ isometry of $\mathrm{CP^2}$. The more complicated three Killing spinors are nontrivial linear combinations of three spinors
\be ({\bf II}) \  \ \ \e_+ \sim e^{-\frac{i}{2} t -\frac{i}{2}y} (e_1^{+-},\e_1^{-+},\e_1^{--}),  \ee
where $\e_1$ depends on $\mathrm{CP^2}$ coordinates nontrivially. They form a triplet under the $SU(3)$ isometry of $\mathrm{CP^2}$. The detail expression is known but not important here.

\end{document}